\documentclass[trackchanges,
amsmath, twocolumn,
]{aastex701}
\usepackage[makeroom]{cancel}
\usepackage{soul}

\begin{document}

\title{Irregular Repeating Tidal Disruption Events due to Diffusive Tides}

\author[orcid=0000-0002-8239-0174, gname='Shu Yan', sname='Lau']{Shu Yan Lau}
\affiliation{eXtreme Gravity Institute, Department of Physics, Montana State University,
Bozeman, Montana 59717, USA}
\email[show]{shuyan.lau@montana.edu}

\author[orcid=0009-0002-8956-7849]{Ethan McKeever}
\email{ethanmckeever@montana.edu}
\affiliation{eXtreme Gravity Institute, Department of Physics, Montana State University,
Bozeman, Montana 59717, USA}

\author[orcid=0000-0002-6011-6190, gname='Hang', sname='Yu']{Hang Yu}
\affiliation{eXtreme Gravity Institute, Department of Physics, Montana State University,
Bozeman, Montana 59717, USA}
\email[show]{hang.yu2@montana.edu}

\begin{abstract}

A repeating partial tidal disruption event (rpTDE) is typically modeled as a star on a bound orbit that is partially disrupted by a massive black hole (MBH) at each pericenter passage. For disruption to occur, the pericenter distance must be close to or within the characteristic tidal radius, such that the tidal field overcomes the star's binding force and triggers mass loss.
However, a binary with a pericenter distance several times the tidal radius can build up its tidal perturbation over multiple orbits via a diffusive process, eventually triggering a nonlinear instability that ejects mass and powers an eruption. This leads to repeated, stochastic disruptions.
In this Letter, we propose that such a mechanism can produce a subclass of rpTDEs with large variations in recurrence time (e.g., J0456-20), which we dub ``diffusive-tide rpTDEs''. We show that diffusive tidal growth can occur for a white dwarf or main-sequence star orbiting a MBH when the pericenter distance is a few times the tidal radius, provided that the orbital period is shorter than the tidal energy dissipation timescale.
These diffusive-tide rpTDEs may account for a significant fraction of the rpTDE population.

\end{abstract}



\section{Introduction} \label{sec:intro}

A tidal disruption event (TDE) occurs when a star passes very close to a massive black hole (MBH) and is disrupted by its tidal field \citep{Rees_1988, Gezari_2021}. 
The pericenter distance, $D_p$, in units of the tidal radius, $R_t$, determines whether the star is fully or partially disrupted. The characteristic $R_t$ is defined as $R_t=(M_\bullet/M_*)^{1/3}R_*$, where $M_*$ and $R_*$ are the mass and radius of the star, while $M_\bullet$ is the mass of the MBH. In a partial TDE, the core of the star remains after a single disruption and may be further disrupted if it remains in a bounded orbit. This leads to repeating partial TDEs (rpTDEs). 

Among the $\sim 200$ TDE candidates observed in different wavelengths \citep{Franz_2025}, a small fraction of them are rpTDEs, with recurring bursts in time intervals of $\mathcal{O}(100)$ days to $\mathcal{O}(10)$ years (e.g., \citealt{Payne_2021, Malyali_2023, Liu_2023, Wevers_2023, Miniutti_2023, Webb_2023, Lin_2024, Sun_2024, Somalwar_2025}). One of the most studied sources is ASASSN-14ko \citep{Payne_2021, Payne_2023}. This source features repeating bursts of about 115~days between flares. One possible physical model describing such events involves an eccentric binary of a star orbiting a MBH with $D_p/R_t \lesssim 1$, such that the tidal force is just strong enough to partially disrupt the star every pericenter passage \citep{Hayasaki_2013, Coughlin_2019, Ryu_2020, Krolik_2020, Nixon_2022, Bandopadhyay_2024, Bandopadhyay_2025}. We refer to these as ``prompt-disruption rpTDEs".

For $D_p$ of a few $R_t$, the weaker tidal force excites the internal normal modes of the star, known as the dynamical tide, after a pericenter passage \citep{Press_1977}. Although the mode amplitude after a single passage is too small to cause disruptions, the excited modes can persist to subsequent pericenter passages and interact with the tidal forcing again.
\cite{Mardling_1995a} first demonstrates that the dynamical tidal interaction in highly eccentric systems can lead to a stochastic evolution of the mode amplitude, resembling a diffusion process (also called `chaotic tide'). The tidal amplitude spreads out as $\sqrt{N_\text{orb}}$, where $N_\text{orb}$ is the number of orbits. This allows the tide to build up and eventually causes the disruption with the release of tidal energy through a catastrophic nonlinear wave breaking \citep{Wu_2018, MacLeod_2022}. We propose that this scenario can lead to a distinct class of rpTDEs, hereafter termed as ``diffusive-tide rpTDEs". Compared to the prompt-disruption rpTDEs, diffusive-tide rpTDEs occur at wider pericenter separations and require multiple orbits to power a single disruption. 

The diffusive behavior is explained by the randomness of the mode phase when the star crosses the pericenter in a highly eccentric orbit with a small $D_p/R_t$ ratio \citep{Ivanov_2004, Vick_2018, Wu_2018}, induced by the tidal backreaction on the orbit. It has recently been shown that orbital decay due to gravitational wave (GW) emission \citep{Ivanov_2007, Lau_2025} and the nonlinear coupling of modes \citep{Yu_2021, Yu_2022, Yu_2026a} can induce similar randomness.
The latter is shown to extend the region of diffusion to less eccentric orbits. 
Eventually, the large tidal amplitude triggers a nonlinear instability, which corresponds to the star exceeding its Roche limit in the circular, synchronous limit \citep{Chandrasekhar_1987, Yu_2026a}. In the highly eccentric limit, the instability further links to the wave breaking hypothesized by \citet{Wu_2018}. The large-scale instability can trigger mass loss from the star, similar to the numerically observed disruption of hot Jupiters \cite{Guillochon_2011}.

The TDEs can originate from the scattering processes in the nuclear cluster that send stars into the disruption region, known as the loss-cone channel \citep{Frank_1976, Merritt_2013}, or the tidal separation of tight binaries by the MBH, known as the Hills mechanism \citep{Hills_1988}. The rates of rpTDEs from these channels have recently been studied by \cite{Pan_2026}.
If these channels can produce eccentric binaries with $D_p/R_t \lesssim 1$ for prompt disruptions to occur \citep{Cufari_2022}, we should also expect them to create systems with greater $D_p/R_t\sim 1-4$ that would experience disruptions due to the diffusive growth of tidal perturbation over multiple orbits.

In this Letter, we propose a subclass of rpTDE produced via the diffusive growth process. In Sec.~\ref{sec:methods}, we describe the formalism to compute the tidal evolution. In Sec.~\ref{ssec:recurr_time}, we study the diffusive growth of tidal energy in a white dwarf (WD)-MBH binary and a main sequence star (MS)-MBH binary, which leads to diffusive-tide rpTDE. In Sec.~\ref{ssec:par_space_orbit}, we provide the orbital parameter space that permits diffusive growth. We then explore the possible fate of the binary after a series of disruptions, depending on the angular momentum loss and the change in radius of the star in Sec.~\ref{ssec:par_space_evolution}. Finally, we summarize and discuss the implications of our results in Sec.~\ref{sec:conclusion}. 

\section{Methods}  \label{sec:methods}


Our model extends \citet{Lau_2025} and involves the coupled evolution of the orbit and the donor star under both tidal interaction \citep{Wu_2018} and non-conservative mass transfer \citep{Sepinsky_2009}. The orbit further evolves under GW radiation \citep{Peters_1963}. For a binary with high eccentricity, the dynamical tide can grow diffusively \citep{Mardling_1995a}, which we track using an iterative mapping following \citet{Vick_2018}. As the tide grows, it eventually becomes unstable and ejects mass from the star \citep{Yu_2026a}. The radius of the star updates following each mass loss episode. Below,  we describe the ingredients entering our model and defer the details to Appendix \ref{app:iterative}.


The tidal deformations of the star are described by the complex mode amplitude $q_a$ from the phase-space expansion following \citet{Schenk_2001}. For binaries with high eccentricities ($e \gtrsim 0.9$), the tidal interactions occur mostly near the pericenter, and $q_a$ behaves as a free oscillator outside this region. This allows us to update $q_a$ and the orbital elements only at each pericenter passage \citep{Mardling_1995b}. The tidal evolution problem now turns into a recursive relation between the quantities of the $(k+1)$th and $k$th orbit (see Appendix~\ref{app:iterative}), involving the mode amplitude $q_a$, orbital energy $E$, and orbital eccentricity $e$. The modes are normalized such that the mode energy $E_{|a|} = E_* |q_a|^2$, where $E_* = GM_*^2/R_*$, and the subscript $|a|$ means that the energy is contributed by a mode indexed by $a$ and its complex conjugate. We shall only consider the dominant $(\ell_a, m_a) = (2,2)$ prograde ($s_a=+$) $f$-mode $(n_a=0)$, its complex conjugate with $(m_{a^\ast},s_{a^\ast})=(-2,-)$, and related leading nonlinear couplings up to four-mode order \citep{Yu_2026a} in this study.

Compared to \cite{Lau_2025}, we further include nonlinear hydrodynamical effects \citep{Weinberg_2012}, which mainly manifest as corrections to the mode frequency $\omega_a$ \citep{Yu_2021, Yu_2023a, Yu_2026a}:
\begin{align}
    \omega^2_\text{eff} =& \omega_a^2\left(1 - \kappa_3 \mathcal{U}_p - \eta_4 |q_a|^2\right), \label{eq:weff}
\end{align}
where $\mathcal{U}_p = \mathcal{U}(D_p)$, with $\mathcal{U}(D) = (M_\bullet/M_*)(R_*/D)^3$ quantifying the instantaneous (normalized) tidal field at separation $D$. The coefficients $\kappa_3$ and $\eta_4$ ($> 0$) are the effective coupling constants at the three- and four-mode level, respectively (see Appendix~\ref{app:iterative} for a discussion). The mode amplitude therefore goes as $q_a \sim e^{-i\omega_\text{eff} t}$ outside pericenter.

The effective frequency provides useful insight into the tidal disruption of the star. The $\kappa_3 \mathcal{U}_p$ term in Eq.~\eqref{eq:weff} dominates the instantaneous tidal field at pericenter when $|q_a|^2$ is small. At small $D_p/R_t$, it causes instability as $\omega_\text{eff}$ turns imaginary. This is interpreted as a critical $D_p/R_t$ for the prompt disruption of the star, as in the circular limit, \citet{Yu_2026a} showed that the same criterion leads to the Roche limit computed in \citet{Chandrasekhar_1987}. Similarly, a large $|q_a|^2$ can cause instability, which may correspond to the tidal wave breaking phenomenon \citep{Wu_2018, MacLeod_2022}. If we approximate $\kappa_3$ and $\eta_4$ using the affine model (see \citealt{Carter_83, Carter_85, Diener_1995, Yu_2026a}, and Appendix~\ref{app:iterative}), we find the prompt disruption and wave breaking to occur at $D_p/R_t = 0.93$ and $|q_a|^2 = E_{|a|}/E_* = 11.3~\%$, respectively, for an $n=1.5$, buoyancy-neutral polytrope.
For the rest of the analysis, we will adopt $D_p/R_t = 1.0$ and $E_{|a|}/E_* = 10~\%$ as the estimated onset of instabilities.

The mode evolution exhibits diffusive behavior if the mode phase, $\omega_\text{eff} P$, varies randomly with a size $\gtrsim 1\,{\rm rad}$ every orbit (e.g., \citealt{Ivanov_2004}), which can be caused by the tidal backreaction \citep{Mardling_1995a, Mardling_1995b, Wu_2018, Vick_2018}, GW backreaction \citep{Ivanov_2007, Lau_2025}, and anharmonicity \citep{Yu_2021, Yu_2022, Yu_2026a} at small $D_p/R_t$ and large $e$. Here, $P$ is the orbital period. For tidal backreaction, the phase perturbation is $\Delta \phi_{\rm BR}=\Delta P \omega_a \simeq (\Delta E_{|a|}/|E|) P \omega_a$, whereas for the nonlinear anharmonicity, $\Delta \phi_{\rm AH} = P \Delta \omega_a \simeq (\Delta E_{|a|} / E_\ast) P \omega_a $ where all the order unity numerical prefactors are dropped. 
The quantity $E$ represents the (time-dependent) orbital energy.
One can show that the two effects add in phase. The ratio is given by 
\begin{equation}
    \frac{\Delta \phi_{\rm AH}}{\Delta \phi_{\rm BR}} \simeq \frac{|E|}{E_\ast} \simeq (1-e) q^{-2/3} \left(\frac{R_t}{D_p}\right),
\end{equation}
where $q = M_*/M_\bullet$.
For the parameter range we consider, with $(1-e)\sim 0.01$ and $q \sim 10^{-5}$, the effect from anharmonicity is the main source of diffusive growth, dominating by a factor of $\sim 10$.

As $E_{|a|}$ reaches the wave-breaking limit at $10~\% E_*$, we reset the mode energy to $0.1~\%E_*$ \citep{Wu_2018}. The energy released can unbind a fractional mass of $|\Delta M_{*}|/M_{*} < 10~\%$ from the star, with $\Delta M_{*}<0$. The mass loss's impact on the orbit (in terms of $\Delta e$) is modeled based on angular momentum balancing:
\begin{widetext}
\begin{align}
    \frac{\Delta J}{J} =& \frac{\Delta M_*}{M_*} + \frac{\Delta M_\bullet}{M_\bullet} -\frac{\Delta M}{2M} + \frac{\Delta e}{2(1+e)}, \label{eq:dJ}\\
    \text{with the parametrization: } \frac{\Delta M_*}{M_*} =& \sigma_1, \;\;\; -\frac{\Delta M_\bullet}{\Delta M_*} = \sigma_2, \;\;\; \frac{\Delta J}{J} = \frac{1-\lambda}{q} \frac{\Delta M}{M},\label{eq:sigma}
\end{align}
\end{widetext}
where $M$ and $\Delta M$ are the binary mass and its change, $J = M_* M_\bullet \sqrt{G D_p(1+e)/M}$ and $\Delta J$ are the orbital angular momentum and its change\footnote{The angular momentum transferred to the mode has not been accounted for in this study. The mode angular momentum contributes only a small fraction to the total angular momentum change, relative to the energy change, by a factor of $(1-e)^{3/2}$ \citep{Ivanov_2004, Vick_2018, Yu_2022}.
However, it can still affect the orbit over a long evolution \citep{Arras_2023}. In addition to the angular momentum transferred to the mode, the tidal torque can contribute to a non-zero precession effect that is not accounted for in this study.}. The parameters $\sigma_1$ ($< 0$), $\sigma_2$ ($ > 0$) and $\lambda$ quantify the fractional mass change of the star, the fraction of the ejected mass transferred to the MBH, and the change in the specific angular momentum (per total binary mass), respectively. The mass transfer results in a change in orbital energy and eccentricity (see Appendix~\ref{app:iterative}).

We remark that the orbital dynamics are almost unaffected by $\sigma_2$ ($ < 1$) for $q \ll 1$. We take $\sigma_2 = \exp{[-1.43(D_p/R_t-2)]}$ for $D_p > 2 R_t$ and $\sigma_2 = 1$ otherwise, following \cite{Lau_2025}. 
The factor, $(1-\lambda)$, quantifies the deviation in the angular momentum loss relative to $[\Delta M/(q M)]J$, which is the loss if the mass is ejected at the center of mass of the star (see Appendix~\ref{app:J_loss}). It can represent the fraction of the specific angular momentum in the ejected mass that is returned to the orbit, as well as its deviation from that carried by a point mass at the center of mass of the star. This model reduces to that in \cite{Lau_2025} if $\lambda = \sigma_2$. In this work, we consider $|\lambda| \leq R_*/D_p \sim q^{1/3}$.

In the following, we consider fiducial WD-MBH and MS-MBH binaries with the stellar and orbit parameters listed in Table~\ref{tab:fiducial} as our fiducial models. To capture the star's structural change after mass loss, we prescribe a phenomenological radius-mass relation. For an MS star, we assume a simple power-law relation, $R_*=R_\odot (M_*/M_\odot)^\alpha$.
For the WD, we employ the fit by \cite{Zalamea_2010} (see their Eq.~(5)), which has a power-law index $\alpha\simeq -0.3$ for low-mass WDs. Note that the post-disruption radius-mass relation may deviate from the unperturbed stellar evolution track due to composition mixing, depending on the stellar age and pericenter separation \citep{Sharma_2024}. For this reason, we treat $\alpha = 0.8$ as the fiducial value for the unperturbed MS but also consider a range of other values to account for the different evolutionary tracks (see Sec.~\ref{ssec:par_space_evolution}).

\begin{deluxetable*}{cccccccccc}
\tablewidth{0pt}
\tablecaption{The parameters of the fiducial binary models. The dynamical frequency is defined as $\omega_{\rm dyn} = \sqrt{GM_*/R_*^3}$. The quantities $Q_a$ and $P_0$ are the overlap integral and initial orbital period. \label{tab:fiducial}}
\tablehead{
\colhead{Binary} & \colhead{$M_*/M_\odot$} & \colhead{$M_\bullet/M_\odot$} & \colhead{$R_* /(10^3 \rm{km})$} & \colhead{$\omega_a /\omega_{\rm dyn}$} & \colhead{$Q_a$} & \colhead{$P_0$}  & \colhead{$D_p/R_t$}& \colhead{$\sigma_1$}
}
\startdata
WD-MBH & $0.5$& $10^{5}$ & $10.6$ & $1.455$ & $0.243$ & $10~\rm hours$ & $2.9$ & $-10^{-3}$ \\
MS-MBH & $0.5$& $10^{5}$ & $4.0\times 10^3$ & $1.0$ & $0.35$ & $45~\rm days$ & $2.0$ & $-0.02$ \\
\enddata
\end{deluxetable*}

\section{Results} \label{sec:results}

\subsection{Recurrence time of the rpTDEs produced by diffusive growth of tide} \label{ssec:recurr_time}

\begin{figure*}[ht!]
\plottwo{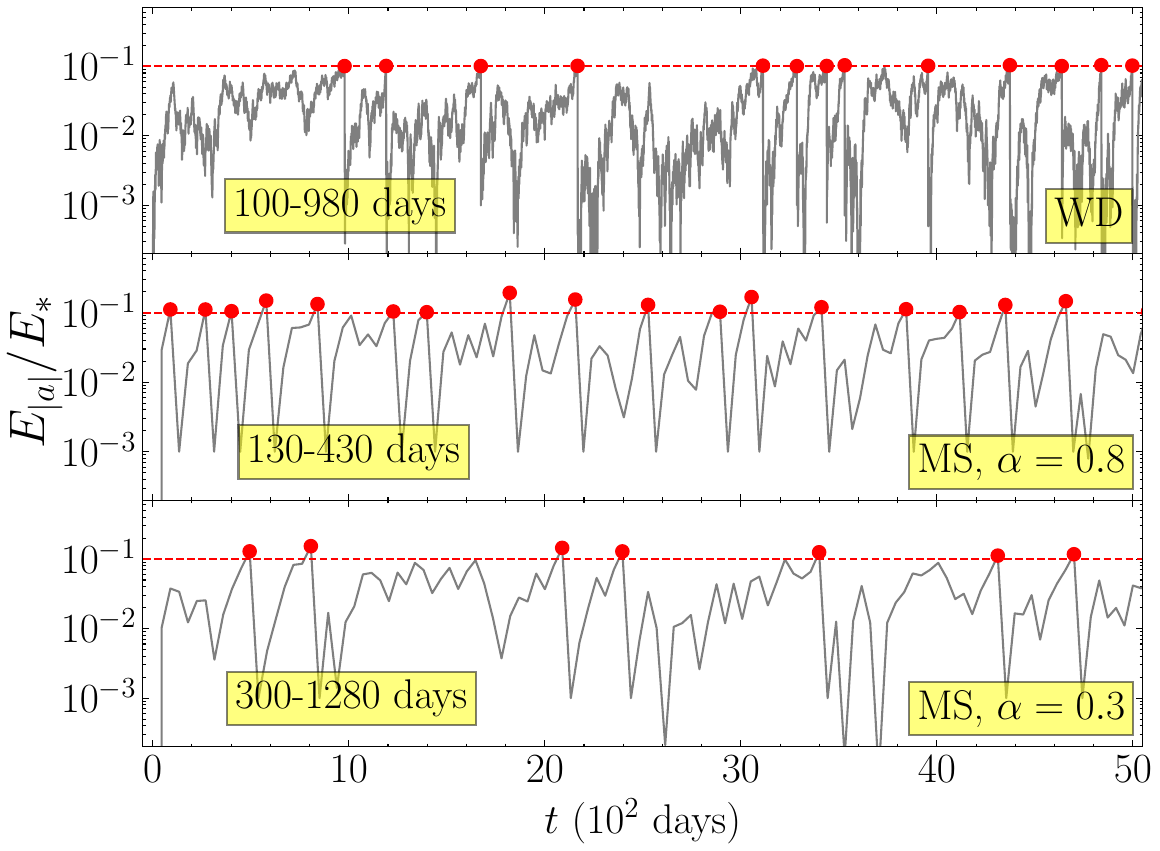}{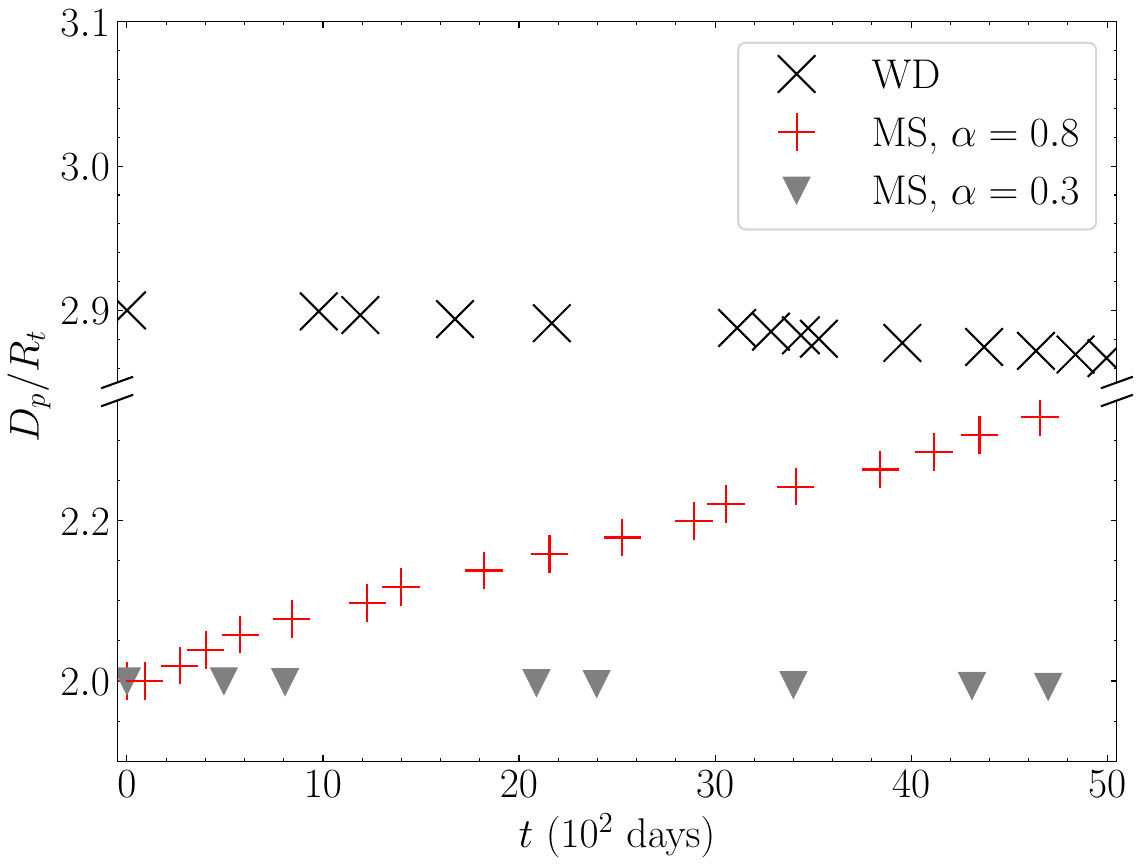}
\caption{Left: The $f$-mode energy evolution of a WD, and two different MSs with $\alpha = 0.3\text{ and } 0.8$, orbiting a MBH. The initial ($D_p/R_t$, $P_0$) are (2.9, 10 hours) for the WD system and (2.0, 45 days) for the MS system. The red horizontal line indicates the wave-breaking threshold, and the red dots show where wave breaking occurs, triggering mass ejection and powering the rpTDEs. The range of the time between mass ejections (within the plotting range) is labeled in each case. The mass loss parameters ($\sigma_1$, $\lambda$) are taken to be ($10^{-3}$, $0$) and ($2\times10^{-2}$, $0$) for the WD and MS respectively.
Right: The evolution of $D_p/R_t$ after each mass ejection.
\label{fig:Emode_vs_time}}
\end{figure*}

Diffusive growth leads to an overall increase in the averaged mode energy with $N_\text{orb}$, eventually causing the partial disruption of the star.
In the left panel of Fig.~\ref{fig:Emode_vs_time}, we show the mode energy evolution of a WD and MS models with different power-law indices $\alpha$. In each case, the mode energy varies stochastically with an increasing trend. As it reaches 10~\% $E_*$, we assume the wave breaks nonlinearly, which induces a fractional mass loss of $\sigma_1$ (Table \ref{tab:fiducial}). The mode energy is reset to $0.1~\% E_*$. The radius is also adjusted according to the $R_*-M_*$ relation as discussed later at the end of Sec. \ref{sec:methods}. This causes a change in $D_p/R_t$ as shown in the right panel of Fig.~\ref{fig:Emode_vs_time}.

\begin{figure*}[ht!]
\plottwo{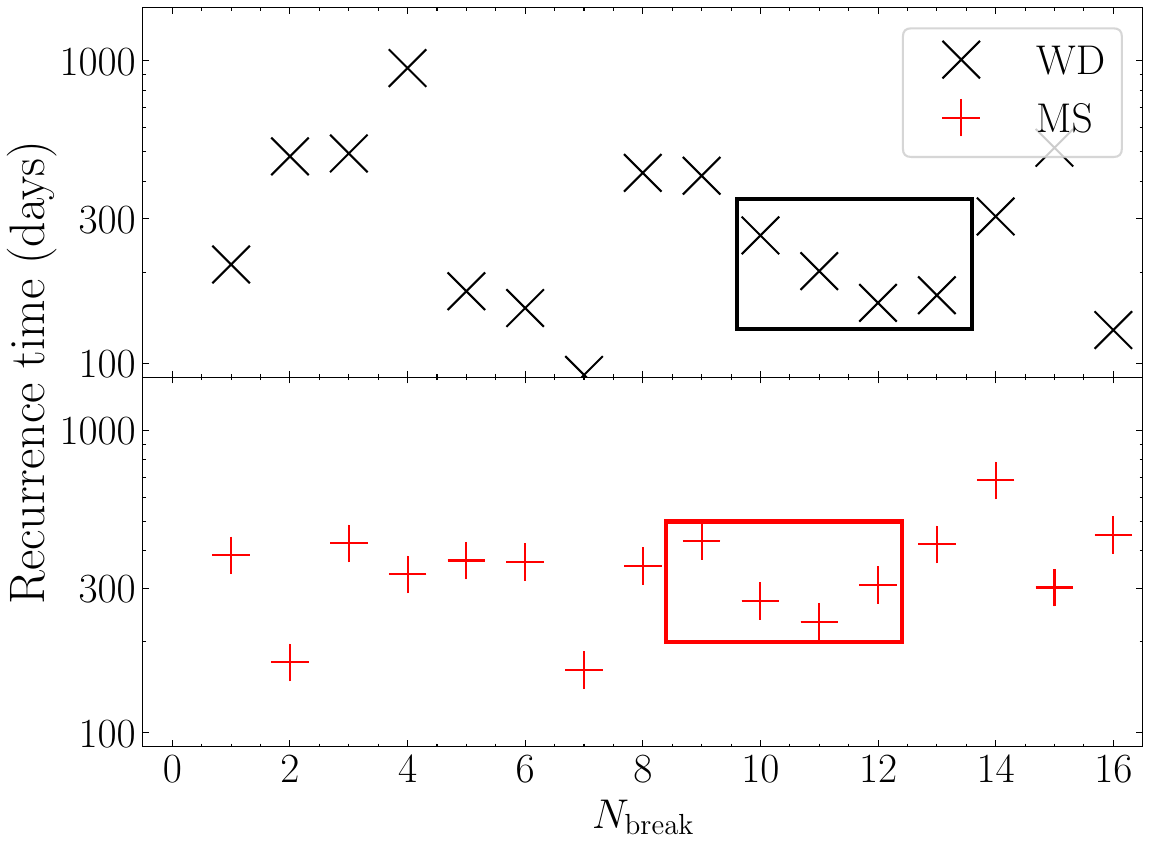}{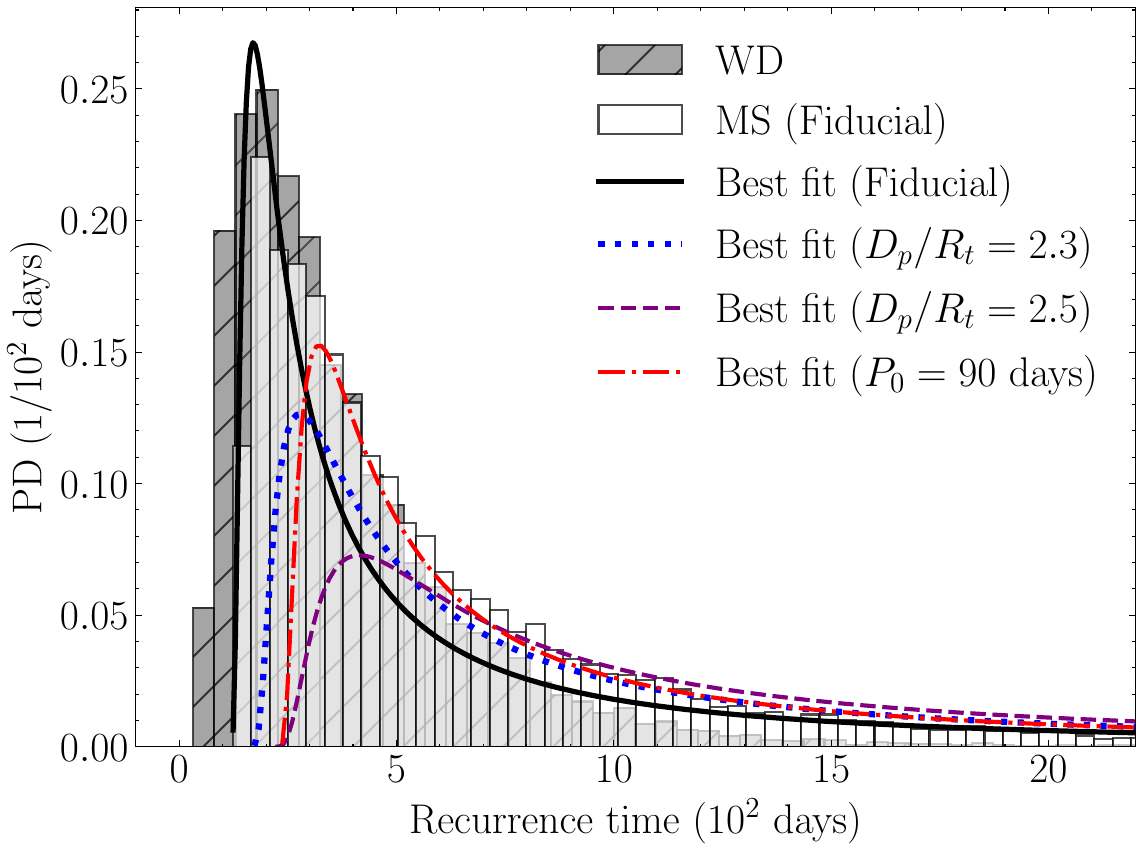}
\caption{Left: The time between successive mass ejection episodes against episode number occurred for the WD system and MS system with $\alpha = 0.8$. The non-monotonic pattern observed in J0456-20 is found in each case, as indicated by the points enclosed by the boxes. 
Right: The probability density of the recurrence time. The solid black line is the best fit with the L\`evy distribution for the fiducial MS system (Table~\ref{tab:fiducial}). Additional lines show fits for models with varied initial orbital parameters. For visual clarity, their corresponding histograms are omitted.}
\label{fig:recurr_time}
\end{figure*}

The stochastic nature of the diffusive tide provides a natural explanation for the large, non-monotonic variations in burst recurrence time, such as those observed in J0456-20 \citep{Liu_2023, Liu_2024}. In the left panel of Fig.~\ref{fig:recurr_time}, we plot the time between mass ejection episodes (via wave breaking) against the episode number for the WD and the MS system with $\alpha = 0.8$. The initial orbital parameters follow those in Fig.~\ref{fig:Emode_vs_time} and Table~\ref{tab:fiducial}.
The right panel shows a histogram of the time between mass ejections over orbits until one of the termination conditions is met: (1) the binary's orbit unbinds, (2) the star loses $90\%$ of the initial mass. 
The samples are created by imposing 900 different initial $q_a$ uniformly spaced with both the real and imaginary part within $[-3\times 10^{-3}, 3\times 10^{-3}]$.
Individual system evolution is highly sensitive to the initial $q_a$ for a given orbit. However, the broader statistical properties of the recurrence time are characterized by the initial orbital parameters.
With the current choice of orbital parameters, the recurrence time shows a distribution that peaks at $\sim 300$ days and gradually declines at longer times.
The heavy-tailed feature can be captured by the L\'evy distribution (see, e.g., \cite{Nolan_2020}). This suggests the long tails are caused by (one-sided) L\'evy flights within a chaotic system \citep{Koren_2007, Manwadkar_2020}. 
Using the fiducial initial parameters, both binaries can reproduce repeating bursts with a statistically averaged recurrence time of $\sim 300$ days and a random variation consistent with that observed in J0456-20. We also illustrate that the location and scale of the distributions depend on the initial parameters by considering various initial $D_p/R_t$ and $P_0$ within the MS system. This implies the possibility of inferring the initial orbit configurations from the observed recurrence times. Therefore, the correlations between the distributions and the binary parameters are worth further study.

The evolution of the star due to mass loss, together with the backreaction on the orbit, results in a change of $D_p/R_t$ as shown in the right panel of Fig.~\ref{fig:Emode_vs_time}. As the star loses mass at the pericenter, the change in $D_p$ is negligible, while $R_t$ changes according to the mass loss as informed by the $R_*-M_*$ relation. For $\alpha>0.33$, stellar density increases following mass loss, causing an increasing $D_p/R_t$ that can suppress the luminosity of the bursts as the mass is less likely to be accreted.
For the case of WD, as well as an MS with $\alpha<0.33$, $D_p/R_t$ decreases upon each mass loss, resulting in an evolution in the opposite direction.

With the parameters in Table~\ref{tab:fiducial}, we find that our models can produce a peak luminosity of $\sim 10^{44}~{\rm erg~s}^{-1}$ if we assume the peak luminosity ($L_\text{peak}$) follows the simple proportionality relation to the peak fall-back rate of the debris ($\dot{M}_{\rm peak}$) \citep{Evans_1989, Ulmer_1999, Bogdanovi_2014}\footnote{More sophisticated models for the accretion physics (e.g., \cite{Metzger_2022, Mummery_2023}) are required to describe the luminosity generation. Here, we treat the simple model only as an order-of-magnitude estimate.}:
\begin{align}
    L_\text{peak} \sim 0.1 \dot{M}_{\rm peak} c^2.
\end{align}
In our model $\dot{M}_{\rm peak} \propto \sigma_2$, which is not constrained by the evolutionary dynamics of the orbit.




\subsection{Orbital parameter space for diffusive growth}\label{ssec:par_space_orbit}

\begin{figure*}[ht!]
\plottwo{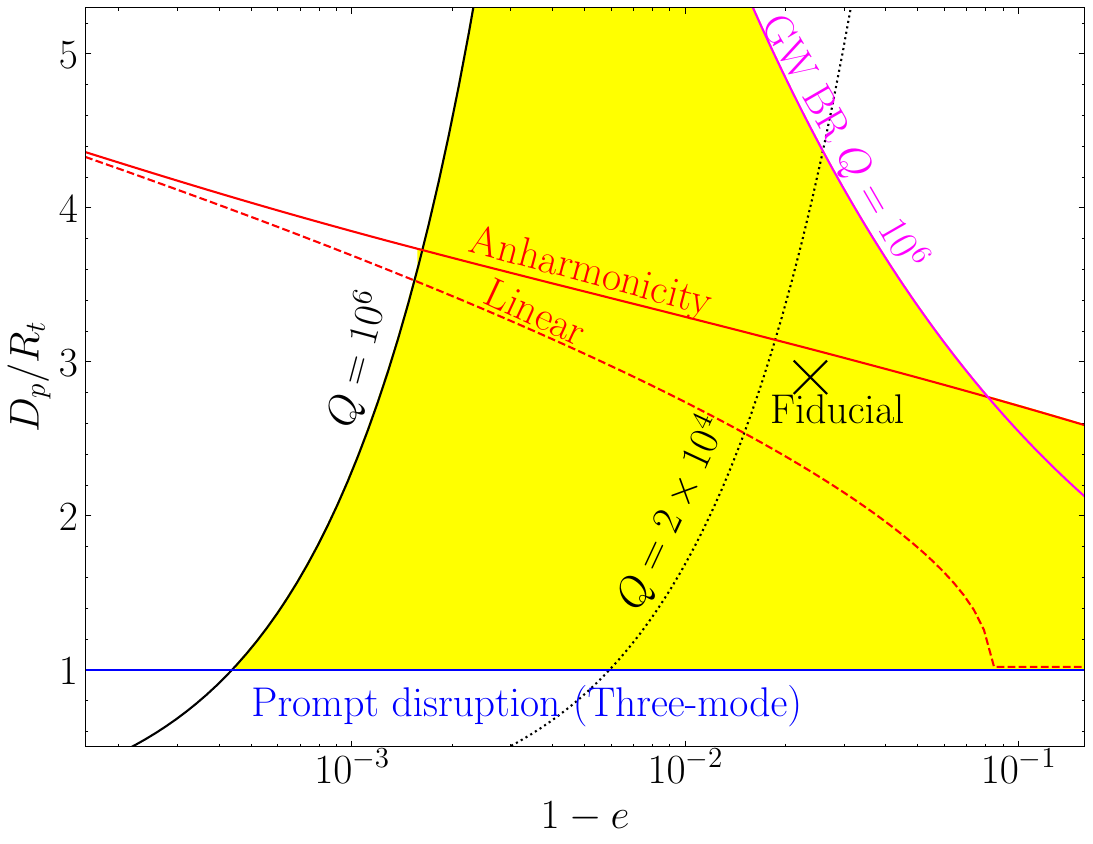}{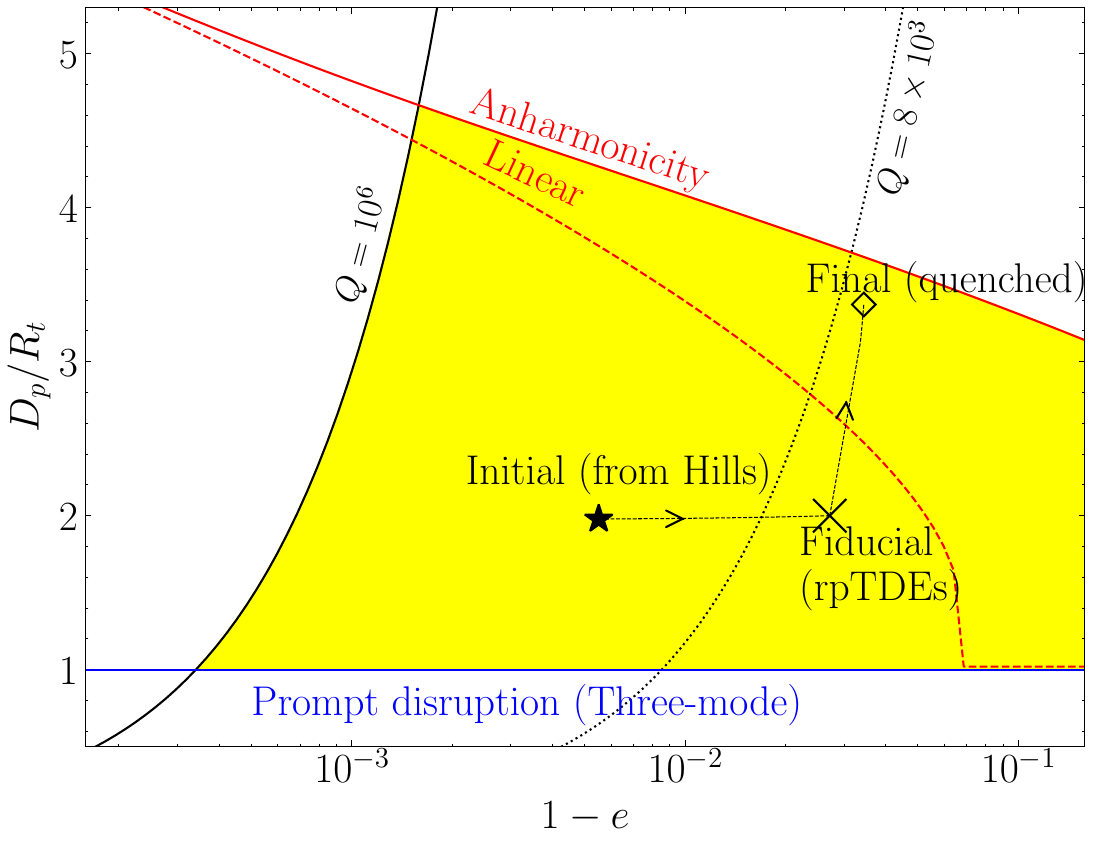}
\caption{
(Left):  The orbital parameter space of WD-MBH system where diffusive-tide rpTDEs are expected is colored in yellow. The black cross indicates the fiducial orbital parameters (see Table~\ref{tab:fiducial}). The black solid line sets the boundary of diffusive growth assuming $Q = 10^6 = \omega_a P_0$, and the dotted line (with $Q = 2\times 10^3$) gives the marginal boundary for the fiducial model. The magenta line represents the threshold for diffusive growth driven by GW backreaction \citep{Lau_2025}. The red solid (dotted) line is the threshold for diffusive growth due to nonlinear anharmonicity (linear tidal backreaction). The blue line corresponds to where prompt disruption is expected.
(Right): Similar plot for an MS-MBH system with $\alpha=0.8$. A potential evolution pathway is shown by the arrowed lines: the MS might be captured on a wide orbit via the Hills mechanism (black star) and circularize, with the diffusive tide suppressed by large dissipation. When the fiducial point (cross) is reached, diffusive tides start to power rpTDEs. The MS becomes denser as it loses mass, eventually quenching the eruptions (diamond).
\label{fig:phase_space_orbit}}
\end{figure*}

Diffusive growth requires a parameter range where the mode phase shifts randomly by $\mathcal{O}(1)$ at pericenter. Figure~\ref{fig:phase_space_orbit} illustrates this $D_p$-$e$ space for the WD-MBH and MS-MBH systems. The black solid and dotted lines denote the threshold $Q = \omega_a P_0$, where $Q$ is the $f$-mode quality factor representing the cycles required for tidal energy to dissipate. Regions below these lines indicate tidal growth exceeding orbital dissipation, enabling diffusive growth.
In both systems, we take $Q = 10^6$ as a reference value. It is a reasonable lower limit of the $f$-mode $Q$ factor for the WD case, as the GW dissipation of the mode energy is of $\sim \mathcal{O}(10^{10})$ \citep{GarcaBerro_2006, Lau_2025}. For the MS-MBH system, however, the value of $Q$ may be much smaller if we take into account parametric instability due to nonlinear mode couplings \citep{Kumar_1996}, especially for medium-low mass MS (0.5-1.2$M_\odot$) with a radiative core. The efficient coupling between the $g$-modes within the core and the $f$-mode can result in $Q$ factors of $\lesssim 10^3$.
To address these uncertainties, the dotted line indicates the marginal $Q$ required for the fiducial model to remain in the diffusive regime. In both systems, diffusive growth persists provided $Q \gtrsim 10^4$.

The backreactions and anharmonicity contribute to the randomness in mode phase. For the WD-MBH system at large $D_p/R_t$, the main contribution comes from GW orbital decay, which drives the $f$-mode resonances with the orbital harmonics \citep{Lau_2025}. This effect is limited by the dissipation of the mode energy between two consecutive resonances, as indicated by the magenta line in the left panel of Fig.~\ref{fig:phase_space_orbit}, assuming $Q = 10^6$. This boundary satisfies
\begin{align}
    Q = \frac{2\pi P_0}{\Delta P_\text{GW}}.
\end{align}
The expression on the right quantifies the number of $f$-mode cycles between successive resonances driven by the GW orbital decay. The secular change in orbital period over a complete orbit due to GW, $\Delta P_\text{GW} = -(3/2) P_0\Delta E_{\rm GW}/E$ (see Eq.~\eqref{eq:Peters_E} of Appendix~\ref{app:iterative}, and \cite{Peters_1963}).

At low eccentricities, the randomness is dominated by the effect of anharmonicity (red solid line). This boundary is estimated by setting the change in the effective mode phase after one kick, $\Delta (\omega_\text{eff} P)$, equal to 1 rad for an $n = 1.5$ polytrope \citep{Vick_2018}. We remark that using an incompressible model reduces the $D_p/R_t$ boundary by $\sim$10~\%. For $D_p/R_t \leq 1.0$, prompt disruptions are expected as discussed in Sec.~\ref{sec:methods}.


Even with efficient dissipations, systems formed with high eccentricities can still migrate towards the region for diffusive growth. 
The right panel of Fig.~\ref{fig:phase_space_orbit} shows a simplified evolution track of such a system initially formed with $e = 0.995$, or an orbital period of 2.3 years (marked by the black star). Employing $Q = 8\times 10^{3}$, we assume the dissipation dominates the tidal evolution initially, such that all the mode energy is dissipated within each orbit. This causes the binary to circularize and evolve toward the fiducial system, at which the diffusive tide can occur. Since $\alpha = 0.8$, $D_p/R_t$ increases due to mass transfer. This eventually makes the system leave the diffusive region (the unfilled diamond marker). Note that wave breaking within the diffusive region effectively dissipates the orbital energy injected into the tide, which also contributes to the circularization and the migration towards the diffusive boundary. Early work in \cite{Mardling_1995b} has already demonstrated such quenching of the diffusive tide due to mode damping.

\subsection{Outcome of the binaries after multiple mass loss episodes}\label{ssec:par_space_evolution}

\begin{figure*}[ht!]
\plottwo{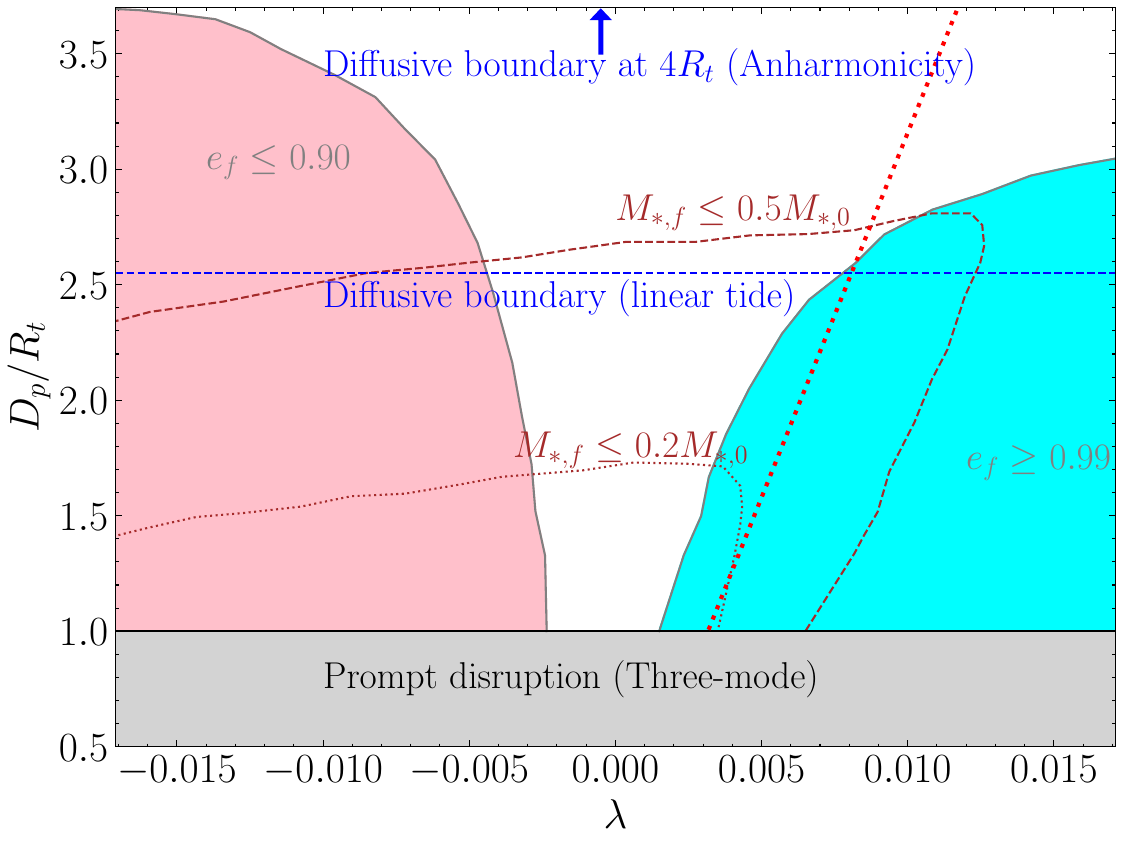}{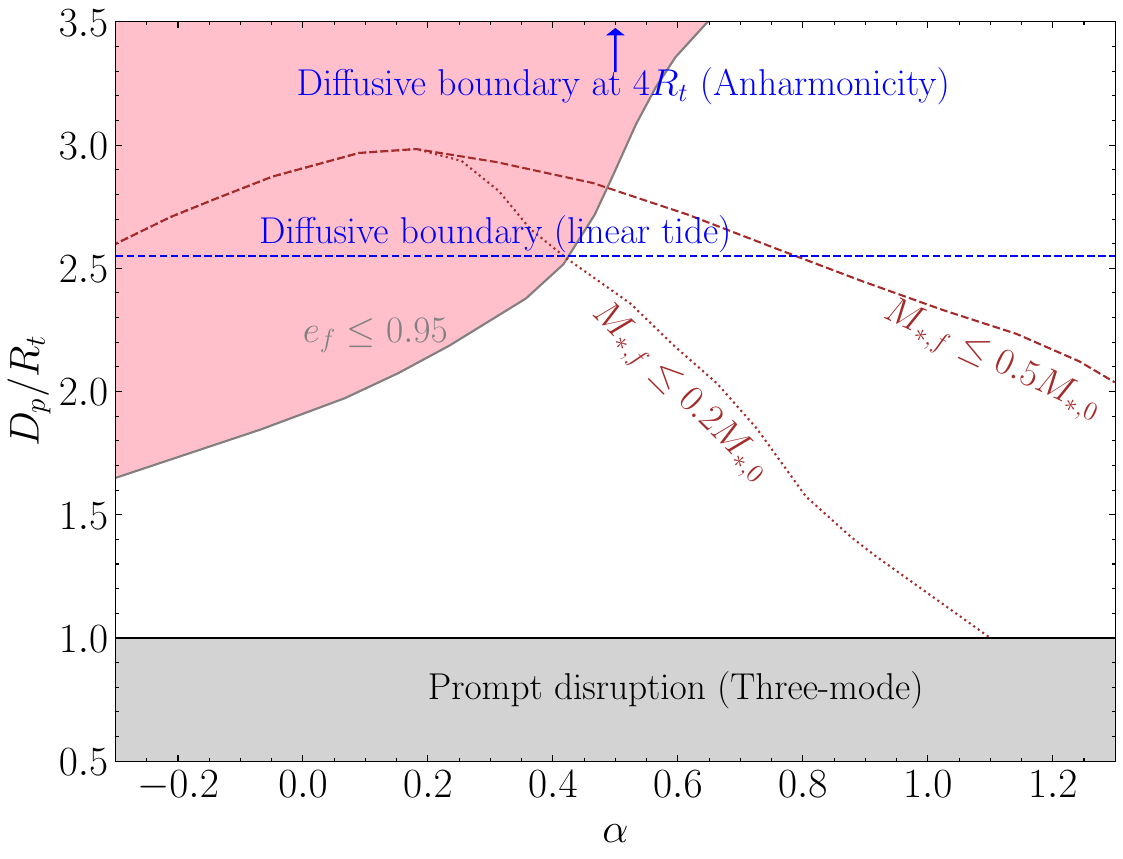}
\caption{
(Left):
Parameter space plot showing the evolution of the MS-MBH binary ($\alpha = 0.8$) for different initial $D_p/R_t$ and $\lambda$. The brown contours indicate the regions where the final mass of the star ($M_{*,f}$) reaches 50~\% and 20~\% of the initial mass ($M_{*, 0}$). 
The pink and cyan regions correspond to orbital circularization and unbinding, respectively.
The red dotted line is the approximate boundary of whether the star or the orbit is disrupted first (Eq.~\eqref{eq:Dp_crit}).
The prompt disruption region is colored gray.
(Right):
Parameter space plot of the evolution for different initial $D_p/R_t$ and $\alpha$, with $\lambda = 0$. 
The pink region now corresponds to $e_f < 0.95$, rather than $0.90$. There is also no region corresponding to $e_f > 0.99$.
\label{fig:phase_space_MS}}
\end{figure*}

Next, we consider the evolution of the binary after a series of mass ejections. The left panel of Fig.~\ref{fig:phase_space_MS} presents the parameter space of the evolution outcome of the MS-MBH binary systems for different ranges of $D_p/R_t$ and $\lambda$, with $|\lambda| \leq R_*/D_p$ (see Appendix~\ref{app:J_loss}).
The binaries are evolved for some orbits until one of the following conditions is met: (1) $e \geq 1$, (2) $M_* \leq 10~\% M_{*, 0}$, (3) the system does not undergo diffusive growth. We set the last criterion as $\langle E_{|a|, k}\rangle < 0.05 k E_{|a|, 0}$ for $k = 5000$, where $\langle E_{|a|, k}\rangle$ is the averaged mode energy up to the $k$th orbit after the previous mass ejection.
For $q \ll 1$, the change in eccentricity (see Eq.~\eqref{eq:de_MT}) can be expanded as
\begin{align}
    \frac{\Delta e}{1-e} = \left(\frac{1+e}{1-e}\right)\left[2\lambda \sigma_1 + q \sigma_1(1-\sigma_2)\right] + \mathcal{O}(q^2). \label{eq:de_MT_expand}
\end{align}
Hence, the mass ejection effect on the orbit is dominated by the $\lambda$ term for $q < |\lambda|$.
A positive $\lambda$ (e.g., correction due to mass ejection at the surface near point) gives a strong repulsive force that can lead to the unbinding of the binary, while a negative $\lambda$ (e.g., near-point mass ejection when the star has a large prograde spin)
yields a circularization effect, as shown in the lower regions of the left panel of Fig.~\ref{fig:phase_space_MS}. Meanwhile, the system can stay at a nearly constant eccentricity if $\lambda$ is small, allowing the star to lose a significant portion of mass. The regions where the star drops below $50~\%$ and $20~\%$ of its initial mass are indicated with the brown dashed line and dotted line, respectively.
Other than the region where $e_f \geq 0.99$ or initial $D_p/R_t \lesssim 1.5$, the iterative map terminates through condition (3), i.e., the evolution becomes periodic. This is mainly caused by the increase in $D_p/R_t$ during mass loss, as the $R_*-M_*$ relation has a power-law index $\alpha = 0.8$. We shall discuss the effect of $\alpha$ on the evolution later in this section.

From Eq.~\eqref{eq:de_MT_expand}, we can approximate (to the leading order in $q$) the boundary separating the region where the orbit expansion effect outweighs the mass stripping, by setting $\Delta e/(1-e) = |\sigma_1|$:
\begin{align}
    D_{p, \text{crit}} = 4 a_0 \lambda, ,\label{eq:Dp_crit}
\end{align}
where $a_0$ is the initial semi-major axis. This is indicated with a red dotted line in Fig.~\ref{fig:phase_space_MS} (left panel).

We also explore the evolution of the WD-MBH binary system using the same parameter range ($|\lambda| \leq R_*/D_p$) and find that the mass loss and orbital circularization are both efficient within the parameter space within the diffusive tide regime. This is because $D_p/R_t$ decreases as the WD loses mass, further enhancing the growth rate of the tidal energy. This causes the system to reach wave breaking at an accelerating rate. The evolution either results in the full disruption of the WD or the circularization of the orbit. 

So far, we have employed a simple $R_*-M_*$ relation to describe the effect of mass loss on the stellar structure of the MS. 
We now quantify the effect of the power-law index $\alpha$ on the evolution of the binary. In the right panel of Fig.~\ref{fig:phase_space_MS}, we show the tendency of the evolution with $\lambda = 0$ for different $D_p/R_t$ and $\alpha$. 
Like in the left panel, the binary is evolved until one of the three conditions specified above is met. Note that since we choose $\lambda = 0$, the mass loss does not increase the eccentricity. The termination sets in only through $M_{*,f} \leq 10~\% M_{*,0}$ or the evolution becomes periodic.
For $\alpha \lesssim 0.33$, mass loss causes $D_p/R_t$ to decrease, thus enhancing the tidal kick amplitude. This increases the eccentricity dissipation, indicated by the pink region. Moreover, the star tends to lose a significant portion of mass ($\geq 80~\%$~$M_{*, 0}$) before the evolution terminates.





\section{Discussions and Conclusion} \label{sec:conclusion}

In this work, we propose a subclass of rpTDE with distinctive timing properties that differ from those described by typical prompt-disruption models, in which the star is disrupted through diffusive tidal growth. This leads to rpTDEs with stochastically varying recurrence times, distinctive from the regularly repeating rpTDEs like ASASSN-14ko. A recently observed rpTDE, J0456-20 \citep{Liu_2024}, that shows large variations in the recurrence time may fall into this subclass (see Figs.~\ref{fig:Emode_vs_time} and~\ref{fig:recurr_time}). These large variations are challenging for previous prompt-disruption models, as they require a 90~\% mass loss of the star in the initial encounter to produce such differences in the orbital period; the amount exceeds what is inferred from the luminosity by about an order of magnitude \citep{Liu_2023, Liu_2024}. We show that such diffusive-tide rpTDEs can occur in a parameter space where $D_p/R_t \lesssim 4$ and $e \lesssim 0.999$ given a tidal quality factor $Q = 10^6$ (see Fig.~\ref{fig:phase_space_orbit}).

We further explore the evolution outcome after multiple mass loss episodes (see Fig.~\ref{fig:phase_space_MS}).
For the MS-MBH system with $\alpha = 0.8$, $D_p/R_t$ tends to increase and can terminate the diffusive growth before the complete disruption of the star unless the initial $D_p/R_t \lesssim 1.7$ (see Fig.~\ref{fig:phase_space_MS}). The mass loss may exert a reaction force onto the orbit that strongly depends on the specific angular momentum loss from the orbit (parametrized by $\lambda$, see Eq.~\eqref{eq:dJ}).
For a highly eccentric orbit, we expect $\lambda\simeq 0$, corresponding to most of the angular momentum carried by the ejected mass being lost from the orbit, as the short pericenter passage would forbid efficient angular momentum feedback. Nonetheless, even a small $\lambda$ can alter the orbit significantly.
For a positive $\lambda$, i.e., a smaller specific angular momentum loss relative to the star's orbit (e.g., mass ejection at a lower orbit), it can unbind the binary system. Otherwise, the diffusive growth eventually stalls due to the increased $D_p/R_t$.
For a WD or a MS with $\alpha \lesssim 0.33$, the mass loss leads to a decrease in $D_p/R_t$, thus enhancing the growth rate of the dynamical tide. The system ends up either having the star fully disrupted or the orbit circularized.

The tidal breakup of a binary, i.e., the Hills mechanism, is a possible formation channel for TDEs \citep{Hills_1988}. \cite{Pan_2026} show that it can be the dominating source of rpTDEs over the loss cone channel \citep{Frank_1976} if a significant fraction of stars ($> 20~\%$) inside the nuclear cluster are in hard binaries.
This mechanism can provide a continuous distribution of $D_p/R_t$ for different progenitor binary periods. Based on Fig.~\ref{fig:phase_space_orbit}, we should expect those with $D_p/R_t \lesssim 1$ to produce prompt mass loss rpTDEs and those with $D_p/R_t\sim 1-4$ to be the potential sources of diffusive-tide rpTDEs featuring large recurrence time fluctuations.

Under the Hills mechanism, the orbital period of the star-MBH system is proportional to the progenitor's period, with a proportionality constant $\simeq q^{-1/2}$ \citep{Cufari_2022}. For the fiducial models (see Table~\ref{tab:fiducial}), it requires progenitor binary periods of less than 3.8 minutes and 0.3 days for the WD and MS system, respectively. 
For the latter, the progenitor's separation is $\sim 2.5\,R_\odot$; MS binaries this compact have been observed \citep{Norton_2011}.
This requirement can be further relaxed if the $f$-mode dissipation is efficient due to, e.g., parametric instabilities or inhomogeneous forcing of p-modes \citep{Kumar_1996}, such that a wider progenitor binary can produce a star-MBH system with a longer orbital period, with mass loss suppressed due to damping of the dynamical tide. As the orbit circularizes to $1-e\gtrsim0.01$, diffusive growth of the f-mode starts to operate and produce rpTDEs (see the evolution track in the right panel of Fig.~\ref{fig:phase_space_orbit}).



Lastly, our model opens up several avenues for further studies. 
Compared to the prompt-disruption model, diffusive-tide rpTDEs may occur at greater distances through the buildup of tide. The extended parameter space can be incorporated further into event-rate studies. 
It is also worth noting that tidal heating can also extend the threshold $D_p/R_t$ by a similar factor \citep{Li_2013, Linial_2023, Yao_2025}. More careful studies integrating both are therefore needed.
Besides, our study considers only an order-of-magnitude estimate of the luminosity. The observed properties of the peak luminosity, e.g., the decay in J0456-20 \citep{Liu_2024}, still require further details of the debris dynamics and luminosity production mechanism.
In addition, the association of the $\kappa_3$ and $\eta_4$ terms to prompt disruption and wave breaking (see Eq.~\eqref{eq:weff}) can be compared with hydrodynamic simulations, potentially allowing an analytic description for the mass loss mechanisms.


\begin{acknowledgments}
We thank Zhen Pan for the suggestions regarding the formation rates. We are also grateful to Rosemary Mardling for the detailed and insightful comments, especially on pointing out the connection between the recurrence times and L\`evy distributions, as well as the distributions' dependence on initial conditions of the orbit.
This work is supported by Montana NASA EPSCoR Research Infrastructure Development under award No. 80NSSC22M0042, NSF award No. PHY-2308415, and CAREER award No. PHY-2541579.
E.M. is supported by NASA FINESST award 80NSSC25K0310.
\end{acknowledgments}

\software{\texttt{NumPy} \citep{harris_2020},  
        \texttt{Numba} \citep{Hunter_2007}, 
          \texttt{Matplotlib} \citep{Hunter_2007}
          }


\appendix

\section{Iterative map} \label{app:iterative}

The main results of this Letter are obtained using an iterative map approximation scheme modified from that in previous studies \citep{Mardling_1995b, Ivanov_2004, Vick_2018}. While it originally contained only the tidal backreaction on the orbit, we extend it to include also the GW backreaction and the effect of anharmonicity from nonlinear mode coupling.

The tidal deformation of the star is described by the Lagrangian displacement vector, $\boldsymbol{\xi}(t, \mathbf{x})$. We adopt the phase-space expansion in terms of the eigenmodes (\citealt{Schenk_2001}):
\begin{align}
    \begin{bmatrix}
        \boldsymbol{\xi} \\
        \dot{\boldsymbol{\xi}}
    \end{bmatrix}
    = \sum_{a} q_a(t)
    \begin{bmatrix}
        \boldsymbol{\xi}_a(\mathbf{x}) \\
        - i \omega_a \boldsymbol{\xi}_a(\mathbf{x})
    \end{bmatrix}, \label{eq:phase_space_expansion}
\end{align}
where $\omega_a$ and $\boldsymbol{\xi}_a$ are the eigenfrequency and eigenfunction of mode $a$. 
The subscript $a$ represents a mode labeled by $(s_a, n_a, \ell_a, m_a)$, representing the sign of the mode frequency, radial, polar, and azimuthal order, respectively.
The mode amplitude $q_a$ follows the equation of motion of a forced oscillator driven by the external tidal force (see, e.g., \citealt{Schenk_2001, Weinberg_2012} for details). For our system, the sum is dominated by the $(s_a, \ell_a, m_a) = (+, 2, 2)$ prograde $f$-mode \citep{Wu_2018}. Note that this mode can couple to other modes when we go beyond the linear level, as described in Sec.~\ref{sec:methods}.

Up to the four-mode coupling level, the mode amplitudes follow a set of first-order nonlinear differential equations \citep{Weinberg_2012, Venumadhav_2013, Weinberg_2016, Yu_2020}:
\begin{align}
    \frac{dq_a}{dt} + i\omega_a q_a = i\omega_a \left(U_a + \sum_{b} U^*_{ab} q_b^* + \sum_{bc} \kappa^*_{abc} q_b^* q_c^*  + \sum_{bcd} \eta^*_{abcd} q_b^* q_c^* q_d^*\right), \label{eq:dq_full}
\end{align}
where $U_a$ and $U^*_{ab}$ are the linear and nonlinear tidal force projected onto the eigenbasis, respectively. The coefficients $\kappa_{abc}$ and $\eta_{abcd}$ are the three-mode and four-mode coupling constants, respectively. The precise definitions of these coefficients can be found in \cite{Weinberg_2012, Venumadhav_2013}. The leading contribution to the dominant mode is given by (cf. eqs. 16, B8, and B10 of \citealt{Yu_2023a})
\begin{align}
    \frac{dq_a}{dt} + i\omega_a q_a = i\omega_a \left\{\sqrt{\frac{3\pi}{10}}I_2  \mathcal{U} + \underbrace{(J_2 + 4 \kappa_2 I_2) \mathcal{U} q_a + [3\eta_2 + 8(\kappa_2^2 + \kappa_{r2}^2)]|q_a|^2 q_a}_{=-(\Delta \omega_a/\omega_a)q_a} \right\} + \text{(other nonlinear terms)}.
    \label{eq:dqa_dominant_nl}
\end{align}
The normalized tidal field strength $\mathcal{U}$ is defined in Eq.~\eqref{eq:weff}.
The tidal overlap integrals, $I_2$ and $J_2$, represent the linear and nonlinear couplings between the $\ell_a = 2$ $f$-modes and the tidal field multipole. The mode coupling coefficients, $\kappa_2$ and $\eta_2$, are the three- and four-mode coupling coefficients between the $\ell_a = 2$ $f$-modes. The coefficient $\kappa_{r2}$ represents the three-mode coupling between the radial mode and two $\ell_a = 2$ $f$-modes. The subscript $2$ in these coefficients means that at least one of the $f$-modes has $m_a = 2$. All the couplings obey the angular momentum selection rule such that the $m_a$ values sum up to zero.
The dominant nonlinear corrections are terms that contain $q_a$, which can be moved to the left-hand side of Eq. (\ref{eq:dqa_dominant_nl}) to become an effective shift of the f-mode frequency, as
\begin{equation}
    \frac{dq_a}{dt} + i(\omega_a + \Delta \omega_a) q_a = i \omega_a\sqrt{\frac{3\pi}{10}}I_a\mathcal{U} + \text{(other nonlinear terms)}.
\end{equation}
To better reveal the hydrodynamical instability that leads to mass transfer, one may instead consider an equivalent second-order system, 
\begin{equation}
    \ddot{Q}_{m_a} + \underbrace{(\omega_a^2+2\Delta\omega_a)}_{=\omega_{\rm eff}^2} Q_{m_a} = 2\omega_a^2\sqrt{\frac{3\pi}{10}}I_a\mathcal{U} + \text{(other nonlinear terms)},
\end{equation}
where $Q_{m_a} = q_{m_a, s_a>0} + q_{m_a,s_a<0}=q_{m_a,s_a>0} + q^{\ast}_{-m_a,s_a>0}$ is the sum of two phase-space modes with the same $m_a$ (and $n_a$ and $l_a$, which are omitted in the subscripts) but opposite signs of eigenfrequencies. Clearly, when $\omega_{\rm eff}^2=\omega_a^2 + 2\Delta \omega_a <0$, exponential growth of the mode occurs. For the large-scale f-mode considered here, this leads to the star exceeding its Roche limit \citep{Yu_2026a}.  

Note that we have not included a linear damping term in Eq.~\eqref{eq:dq_full} as in \cite{Weinberg_2012}. At linear level, the $f$-mode damping time is $\gtrsim 10^{4}$ years \citep{Ray_1987, Kumar_1996} for low-mass (convective) MSs, corresponding to a $Q \gtrsim \mathcal{O}(10^7)$. Meanwhile the WD has a $Q$ factor $\gtrsim \mathcal{O}(10^{10})$ from GW dissipation \citep{GarcaBerro_2006, Lau_2025}. Both of them are negligible in the evolution timescale. On the other hand, the nonlinear damping is captured in the wave-breaking process and takes effect only when the mode amplitude is large enough such that $\omega_{\rm eff}^2<0$; see discussions following Eq.~\eqref{eq:weff}.
We ignore weakly nonlinear dissipation (from f-modes nonlinearly interacting with other modes; \citealt{Kumar_1996, Weinberg_2012, Yu_2022}), but have accounted for dissipations beyond the linear prediction in the phase diagram (Fig. \ref{fig:phase_space_orbit}) by including values of $Q$ much smaller than what the linear theory can provide.

In the iterative map, the recurrence relation between the $(k+1)$th and $k$th orbit is given by \citep{Vick_2018}:
\begin{align}
    \bar q_{a,k+1} =& \left[\bar q_{a,k} + \Delta q_{a}(z_\text{eff})\right] \exp{(- i \omega_\text{eff} P_{k+1})}, \label{eq:map_mod}\\
    E_{k+1} =& E_{k} + \Delta E_\text{T} + \Delta E_\text{GW} + \Delta E_\text{MT}, \label{eq:map_E}\\
    e_{k+1} =& e_{k} + \Delta e_\text{T} + \Delta e_\text{GW} + \Delta e_\text{MT}. \label{eq:map_e}
\end{align}
The subscript $k$ denotes the $k$th orbit. The phase-shifted amplitude is $\bar q_{a,k} = q_{a,k}\exp(-i\omega_a P_k/2)$, where $\omega_a$ is the linear mode frequency and $\Delta q_a$ is the tidal kick at pericenter. The effective quantities $z_{\mathrm{eff}}$ and $\omega_{\mathrm{eff}}$ include nonlinear hydrodynamics. The orbital energy, eccentricity, and period are denoted by $E_k$, $e_k$, and $P_k$, with per-orbit changes $\Delta E$ and $\Delta e$. Subscripts T, GW, and MT denote tidal backreaction, gravitational-wave backreaction, and mass transfer.

In this paper, we only consider the $(\ell_a, m_a) = (2,2)$ prograde $f$-mode, which is the dominant contribution to the tide. As shown by \cite{Lai_1997}, $\Delta q_{a}$ for this mode can be expressed as an asymptotic series
\begin{align}
    \Delta q_{a}(z) \propto z^{5/2} \exp{(-2z/3)}\left(1 - \frac{\sqrt{\pi}}{4\sqrt{z}} + \mathcal{O}(z^{-1})\right).
\end{align}
where $z = \omega_a \sqrt{2D_p^3/(GM)}$. We then define $z_\text{eff} = \omega_\text{eff} \sqrt{2D_p^3/(GM)}$.

The leading correction to the tidal kick enters at the three-mode level due to the large $\mathcal{U}_p$ at the pericenter passage except when $\eta_4|q_a|^2$ approaches unity.
In contrast, the main correction to the mode phase $\omega_\text{eff} P_k$ enters at the free-propagation stage away from the pericenter. Since $\mathcal{U}$ falls off as $1/D^{3}$ and the pericenter passage timescale is $\sim D_p^{3/2}$, the $\kappa_3$ contribution to the mode phase is negligible here. Effectively, the amplitude dependence of $\omega_\text{eff}$ makes the star an anharmonic oscillator. For these reasons, we only implement the $\kappa_3$ correction to $\Delta q_{a}(z_\text{eff})$ and the $\eta_4$ correction to $\omega_\text{eff} P_k$ in our code. 

The coefficients $\kappa_3$, $\eta_4$ depend on the stellar model and generally require numerical integration to determine their values. However, for polytropic stellar models, they can be approximated with the affine model introduced in \cite{Carter_83, Carter_85} to give an analytic expression (see Appendix A of \citealt{Yu_2026a})\footnote{Note that \cite{Yu_2026a} uses the configuration space expansion containing only the positive frequency modes, with different normalizations of the eigenfunctions. The effective coupling constants in this work are defined in the phase-space expansion and contain contributions from both positive and negative frequency modes.}.:
\begin{align}
    \kappa_3 =& \frac{65}{28}\left(5-n\right)\frac{\mathcal{M}}{M_*R_*^2}, \label{eq:kappa3}\\
    \eta_4 =& \frac{(5-n) (300 \Gamma -253)}{98 (3\Gamma -4)}, \label{eq:eta4}
\end{align}
where $n$ is the polytropic index, $\Gamma$ is the adiabatic index, and $\mathcal{M} = \sum_i \int x_i x_i dM/3$ is one-third the trace of the mass quadrupole moment of the star in the spherically symmetric configuration. 
Using the hydrostatic equation, one finds $\mathcal{M} = 0.2 M_* R_*^2$ for $n = 0$ and $\mathcal{M} = 0.102 M_* R_*^2$ for $n = 1.5$.
For simplicity, we use $\Gamma = 1+1/n$ in the following when evaluating $\eta_4$.

In this work, the nonlinear interactions mainly come from the coupling between $\ell_a=2$ $f$-modes with different $m_a$, and the fundamental radial mode. Together with the angular momentum selection rule, we consider the coupling between modes with $m_a = (2, 0, -2)$ at the three-mode level, and $m_a = (2,-2,2,-2)$ at the four-mode level (including all permutations). Note that when the $m_a=2$ mode first corrects the $m_a=0$ mode at the three-mode level, the back-reaction of the $m_a=0$ mode on the $m_a=2$ mode via again the three-mode interaction effectively creates a four-mode interaction that enters at the same order as the direct four-mode interaction.
This leads us to the effective corrections in Eq.~\eqref{eq:weff}. The details of deriving the effective coupling constants $\kappa_3$ and $\eta_4$ within the affine model can be found in Appendix~A of \cite{Yu_2026a}.


The change in orbital energy and eccentricity per orbit from the tidal and GW backreaction in the $k$th orbit are \citep{Vick_2018, Peters_1963}\footnote{Formally, in the energy conservation equation, one should also account for energy due to three- and four-mode interactions. However, as fig. 5 of \cite{Yu_2023a} shows, most of the nonlinear effects are captured as long as one corrects the mode amplitude while still using the linear form of mode energy. The energy stored in the nonlinear interactions is subdominant.}
\begin{align}
    \Delta E_\text{T} =& -\frac{G M_{*,k}^2}{R_{*,k}}\left(|q_{a,k+1}|^2 - |q_{a,k}|^2\right),\\
    \Delta e_\text{T} =& -\frac{\Delta E_\text{T}}{E_k}\frac{1-e_k^2}{2 e_k},\\
    \Delta E_\text{GW} =& \frac{128\pi}{5}\frac{M_*M_\bullet}{M^2} E_k \left(\frac{GM}{c^2 a_k} \right)^{5/2} \frac{1+\frac{73}{24}e_k^2 + \frac{37}{96}e_k^4}{(1-e_k^2)^{7/2}},\label{eq:Peters_E}\\
    \Delta e_{\text{GW}} =& -\frac{608 \pi}{15}\frac{M_*M_\bullet}{M^2} \left(\frac{G M}{c^2 a_k} \right)^{5/2} \frac{e_k+\frac{121}{304}e_k^3}{(1-e_k^2)^{5/2}} \label{eq:Peters_e},
\end{align}
where $a_k = -GM_*M_\bullet/(2E_k)$.
Using Eqs.~\eqref{eq:dJ} and \eqref{eq:sigma}, the mass loss contributions $\Delta e_\text{MT}$ and $\Delta e_\text{MT}$ are also explicitly written as:
\begin{align}
    \frac{\Delta e_\text{MT}}{1-e_k} =& \left(\frac{1+e_k}{1-e_k}\right)\left\{2\left[1-(1-\lambda)\frac{1}{1+q}\right]\sigma_1 - 2 q \sigma_1\sigma_2 + \frac{q}{1+q}\sigma_1(\sigma_2-1)\right\}, \label{eq:de_MT}\\
    \frac{\Delta E_\text{MT}}{E_k} =& -\frac{\Delta e_\text{MT}}{1-e_k} + \sigma_1(1-\sigma_2 q). \label{eq:dE_MT}
\end{align}

\section{Angular momentum loss during nonconservative mass transfer} \label{app:J_loss}

The angular momentum carried by the ejected mass during nonconservative mass transfer is parametrized by $(1-\lambda)$ in Eqs.~\eqref{eq:de_MT} and \eqref{eq:dE_MT}.
If the mass is ejected at the same orbit as the center of mass of the star and escapes the system, the angular momentum change of the orbit is given by \citep{Huang_1963, MacLeod_2018, Lau_2025}:
\begin{align}
    \frac{\Delta J}{J} = \frac{1}{q}\frac{\Delta M}{M}.
\end{align}
Meanwhile, the point mass ejected at the surface has a specific angular momentum of $\frac{1-\lambda}{q}|\frac{\Delta M}{M}|$, where $|\lambda| \leq R_*/D_p < 1$, and $\lambda = \pm R_*/D_p$ corresponds to the point mass closest ($+$) or furthest ($-$) to the MBH. If we assume $D_p$ is below a few $R_t$, we have $\sim q^{1/3}$.
For a rotating star with spin rate up to the Keplerian frequency, $\lambda$ is further modified by up to $\sim \sqrt{(M_*/M_\bullet)(D_p/R_*)} \sim q^{1/3}$.

If an accretion disk is formed by the mass transferred towards the MBH, part of the angular momentum can be returned to the orbit via tidal interactions \citep{Priedhorsky_1988}. This can also contribute a positive value to $\lambda$. In the case of rpTDE from eccentric binaries, it is unlikely for the accreted mass to efficiently transfer the angular momentum back to the orbit via tidal torque. Hence, we have not considered the range of $\lambda$ greater than $R_*/D_p$.


\clearpage

\bibliography{references}{}
\bibliographystyle{aasjournalv7}



\end{document}